\documentclass[twocolumn,english,prl,showpacs]{revtex4}
\usepackage[T1]{fontenc}
\usepackage[latin9]{inputenc}
\usepackage{amsmath}
\usepackage{amssymb}
\usepackage{graphicx}
\usepackage{esint}

\makeatletter
\@ifundefined{textcolor}{}
{%
 \definecolor{BLACK}{gray}{0}
 \definecolor{WHITE}{gray}{1}
 \definecolor{RED}{rgb}{1,0,0}
 \definecolor{GREEN}{rgb}{0,1,0}
 \definecolor{BLUE}{rgb}{0,0,1}
 \definecolor{CYAN}{cmyk}{1,0,0,0}
 \definecolor{MAGENTA}{cmyk}{0,1,0,0}
 \definecolor{YELLOW}{cmyk}{0,0,1,0}
 }

\makeatother

\usepackage{babel}
\begin{document}

\title{Topological phases in adiabatic and non-adiabatic driven systems}

\author{A. G\'{o}mez-Le\'{o}n}

\affiliation{Instituto de Ciencia de Materiales de Madrid (ICMM-CSIC), Cantoblanco,
28049 Madrid, Spain.}

\author{G. Platero}

\affiliation{Instituto de Ciencia de Materiales de Madrid (ICMM-CSIC), Cantoblanco,
28049 Madrid, Spain.}

\date{\today}
\begin{abstract}
In this work we study the geometrical and topological properties of
non-equilibrium quantum systems driven by ac fields. We consider two
tunnel coupled spin qubits driven by either spatially homogeneous
or inhomogeneous ac fields. Our analysis is an extension of the classical
model introduced by Berry with the addition of the spatial degree
of freedom. We calculate the Berry and Aharonov-Anandan geometric
phases, and demonstrate the influence of the different field parameters
in the geometric properties. We also discuss the topological properties
associated with the different driving regimes, and show that by tuning
the different parameters one can induce topological phase transitions,
even in the non-adiabatic regime.
\end{abstract}
\maketitle

\paragraph*{Introduction:}

Geometric phases in quantum physics is a fundamental issue which has
been addressed in the last decades. Since Berry's discovery of geometric
phases in quantum systems \citep{Berry84}, and their further applications
to different branches of physics (e.g. condensed matter and optics),
geometry, topology and nature seem to be mixed at a highly complex
level. The characterization of different physical properties in purely
geometrical and topological terms, such as quantized transport\citep{Thouless1983}
and electric polarization\citep{King-Smith1993}, has lead to the
understanding of some undergoing processes governing the physical
features of these systems \citep{Xiao2010}. An important characteristic
of geometrical and topological properties in physical systems, arises
in their strength against different perturbations. Some examples of
those are backscattering in edge states and fault tolerant quantum
computation\citep{Konig2007,Kitaev2003,Dennis2002,Zanardi1999}.

Theory of Principal Fiber Bundles (PFB) has also played an important
role in quantum mechanics due to its suitability to describe the underlying
geometrical and topological properties\citep{Yang1975,Bohm03}. Its
relation with condensed matter was pointed out by Simon \citep{Simon1983},
and the existence of an universal PFB \citep{Anandan1987,Bohm1993},
which is a purely geometric object, shows that non-adiabatic processes
can be used for practical purposes in a similar way than their adiabatic
partners.

The interplay between geometry and ac-fields is a very promising field
with a huge activity in the last years. The tuning of geometrical
properties by the application of ac-fields offers a very interesting
scenario, where non-equilibrium systems show quantum phase transitions,
and the possibility to create topologically protected states\citep{Lindner2010,Kitagawa2010,Bastidas2012,Tomka2012}.

As Berry demonstrated, the phase acquired during the cyclic evolution
of a general quantum system can be expressed as a combination of two
terms, a dynamical phase $\gamma_{D}$ and a geometric phase $\gamma_{G}$,
being the latter obtained due to the parallel transportation of the
vector state through the base manifold\citep{Anandan87}. The parallel
transportation is obtained by means of the 1-form connection $A^{n}:=i\langle n|d|n\rangle$
($d$ is the exterior derivative operator, and $|n\rangle$ is the
instantaneous eigenstate), and the geometric phase acquired is given
by $\gamma_{G}^{n}=\int_{C}A^{n}$, that only depends on the path
followed in the parameter space.

These geometric quantities describe the evolution of the system as
the external parameters are varied, but other quantities, such us
Chern numbers (topological invariants) can also be defined through
the 1-form $A^{n}$, characterizing the topological properties of
the whole base manifold.

\begin{figure}[h]
\includegraphics[scale=0.1]{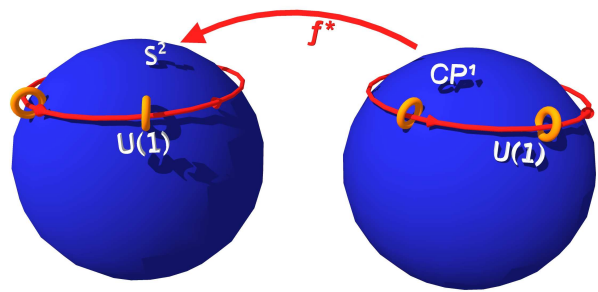}

\caption{\label{fig:Schematic}Schematic figure for the relation between adiabatic
and non-adiabatic regimes. The function $f^{*}$ maps curves $\mathcal{C}$
from the Bloch sphere (projective Hilbert $\mathbb{C}P^{N-1}$) into
curves $C$ in the parameter space $\mathcal{S}^{2}=\left\{ \theta,\varphi\right\} $.
The $U\left(1\right)$ fibers attached to each point due to the phase
invariance of quantum mechanics are represented by small circles.
We consider $\mathbb{C}P^{N-1}$ with $N=2$ for the schematic figure
(i.e.. $\mathbb{C}P^{1}$ which is the Bloch sphere for a single qubit).}
\end{figure}

In the present work we consider an extension of the classical model
considered by Berry\citep{Berry84}, including the spatial degree
of freedom, inhomogeneous magnetic fields and non adiabatic evolution.
The analytical results show that the interplay between the spin and
spatial degree of freedom is remarkably relevant when the ac-field
varies between different sites. We calculate the geometric phase and
Chern number for both adiabatic and non adiabatic evolution. This
model can be realized in different experimental setups, and the results
can be easily extended to larger size systems such as atomic arrays
with a pseudo-spin degree of freedom. We demonstrate that adiabatic
evolution for the ac-field allows tunable geometric phases by varying
the external field parameters: $B$ (intensity of the external field)
and $\phi$ (phase difference between sites). We show that the variation
of the phase difference $\phi$, which introduces the field anisotropy,
strongly modifies the Berry phase and the topological invariants.
Also that non-adiabatic processes increase the tuning possibilities
with the addition of new topological phases, which only appear out
of the adiabatic regime.

\paragraph*{Model:}

The system contains a qubit tunneling between two sites $\left(L,R\right)$
in presence of and ac-field:
\begin{eqnarray}
H\left(t\right) & = & \sum_{i=L,R}B\cos\left(\theta\right)S_{z}^{i}+\sum_{\sigma,i\neq j}t_{LR}c_{\sigma,i}^{\dagger}c_{\sigma,j}+\label{eq:Hamiltonian1}\\
 &  & \sum_{i=L,R}B\sin\left(\theta\right)\left[\cos\left(\varphi_{i}\left(t\right)\right)S_{x}^{i}+\sin\left(\varphi_{i}\left(t\right)\right)S_{y}^{i}\right].\nonumber 
\end{eqnarray}
 The coupling between sites is given by $t_{LR}$, $B_{z}$ is the
static field which splits the qubit levels, and the ac-field $\vec{B}_{ac}\left(t\right)$
couples the different states of the qubit (perpendicular to $B_{z}$).
In this Hamiltonian, the field has been parametrized according to
the angles of a 2-sphere $\mathcal{S}^{2}=\left\{ \theta,\varphi\right\} $,
being $B_{z}\equiv B\cos\left(\theta\right)$, $B_{ac}^{x}\equiv B\sin\left(\theta\right)\cos\left(\varphi_{i}\left(t\right)\right)$
and $B_{ac}^{y}\equiv B\sin\left(\theta\right)\sin\left(\varphi_{i}\left(t\right)\right)$.
This is a very appropriate parameter space for a circularly polarized
ac field with time dependence $\varphi_{i}\left(t\right)=\Omega t+\phi_{i}$,
which is the case considered in this work. For simplicity, we consider
symmetrical $B_{z}$ and $B_{ac}\left(t\right)$ at each site.

For this Hamiltonian we calculate the geometric and topological properties
by means of the connection 1-form $A^{n}$. We consider both, the
adiabatic and non-adiabatic time evolution for phase differences $\phi=\phi_{L}-\phi_{R}\in\left\{ 0,\pi\right\} $,
i.e. magnetic fields in phase or in phase opposition respectively.
We show that in the last case, interesting topological properties
arise. Furthermore, we calculate the phase diagram demonstrating that
non-adiabatic evolution leads to the appearance of new topological
phases.

\paragraph*{Adiabatic case:}

\begin{figure}[h]
\includegraphics[scale=0.75]{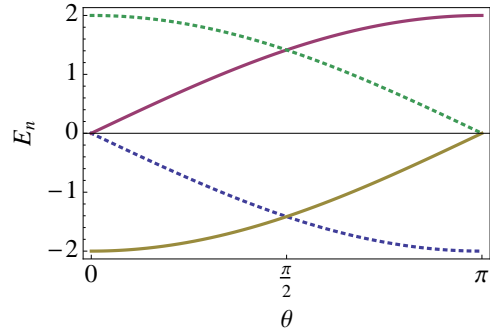}

\caption{\label{fig:Instantaneous-energies1}Instantaneous energies $E_{n}\left(\pi\right)$
vs $\theta$ for $t_{LR}=1$, $\phi=\pi$ and $B=2$ (dashed lines:
$E_{\pm,\pm}\left(\pi\right)$ and continuous lines: $E_{\mp,\pm}\left(\pi\right)$
). }
\end{figure}

The adiabatic case considers slow time evolution. This allows to define
instantaneous eigenstates and neglect the transitions to other energy
levels. The instantaneous energies $E_{n}\left(\phi\right)$ can be
characterized by two indexes after the diagonalization $n\equiv\left(m_{1}=\pm1,m_{2}=\pm1\right)$:
\begin{eqnarray}
E_{n}\left(0\right) & = & -\frac{B}{2}\left(m_{1}+m_{2}\lambda\right)\label{eq:Energy1}\\
E_{n}\left(\pi\right) & = & -\frac{m_{1}B}{2}\sqrt{1+\lambda^{2}-2m_{2}\lambda\cos\left(\theta\right)},\nonumber 
\end{eqnarray}
being $\lambda=2t_{LR}/B$. We can see that both instantaneous energies
are $\varphi$ independent. The case $\phi=0$ reflects the classical
result obtained by Berry (i.e.. without $\theta$ dependence), with
an extra splitting due to the tunneling $t_{LR}$. However, the case
$\phi=\pi$ shows a $\theta$ dependence which has not been obtained
previously. That is, the instantaneous energies now depend on the
ratio between the ac-field and the Zeeman splitting through $\theta$
(Fig.\ref{fig:Instantaneous-energies1}), and not just on the total
intensity $B=|\vec{B}|=\sqrt{B_{ac}^{2}+B_{z}^{2}}$.

In the calculation of the instantaneous eigenstates, due to the parameter
space which is a 2-sphere, we need to consider two different charts.
This is due to the fact that the surface of a sphere cannot be directly
mapped to a plane. The transformation of the instantaneous eigenstates
between different charts is given by phase factors\citep{Bohm03}.
This is a signature of non-trivial geometrical and topological properties.
However, we shall not worry about this, because we are interested
in the geometric phase and the curvature tensor $F^{n}=dA^{n}$ (needed
for the calculation of the Chern number), which are globally defined
for the case of Abelian theories.

\begin{figure}[h]
\includegraphics[scale=0.75]{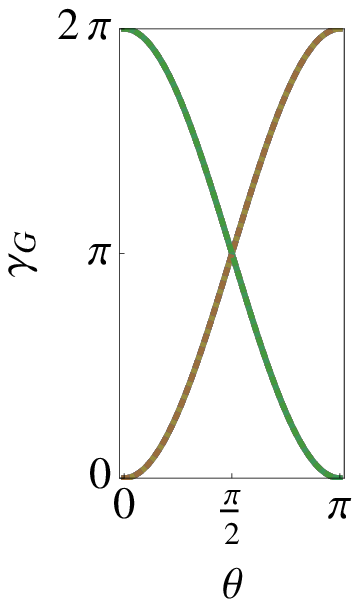}\includegraphics[scale=0.75]{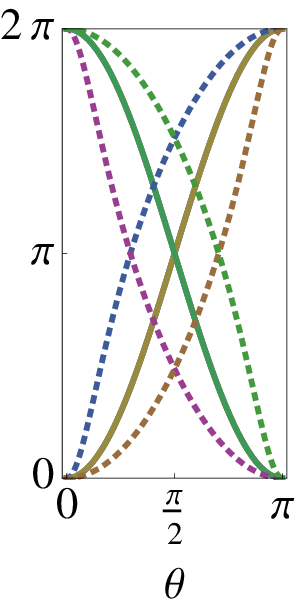}\includegraphics[scale=0.75]{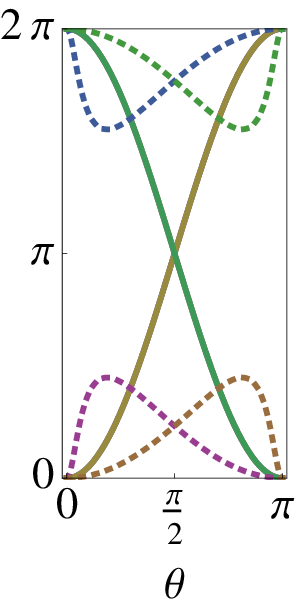}

\caption{\label{fig:Geometric-phase-variation}Geometric phase in the adiabatic
limit $\gamma_{G}$ vs $\theta$ for different $\lambda$ ($\lambda=0$
(left), $\lambda=0.6$ (center) and $\lambda=1.2$ (right)). Dashed
(continuous) lines: $\phi=\pi$ $\left(\phi=0\right)$. Note that
as $\lambda=2t_{LR}/B$ is increased, a gap appears for the case $\phi=\pi$.
The value at which the topological phase transition occurs is exactly
where the gap opens ($\lambda=1$).}
\end{figure}

In order to obtain the geometric phase, we calculate, using the instantaneous
eigenvectors, the connection 1-form $\mathbf{A}^{n}=\left(A_{\varphi}^{n},A_{\theta}^{n}\right)$.
The result shows that the phase difference strongly modifies the geometric
phase (Fig.\ref{fig:Geometric-phase-variation}), which is $t_{LR}$
dependent for the case of $\phi=\pi$:
\begin{eqnarray}
\gamma_{G}^{n}\left(\phi=0\right) & = & \pi\left(1-m_{1}\cos\left(\theta\right)\right)\label{eq:BerryphaseAd}\\
\gamma_{G}^{n}\left(\phi=\pi\right) & = & \pi\frac{m_{1}\left(\lambda m_{2}-\cos\left(\theta\right)\right)+f_{m_{2}}\left(\lambda,\theta\right)}{f_{m_{2}}\left(\lambda,\theta\right)},\label{eq:BerryphaseAd2}
\end{eqnarray}
being $f_{m_{2}}\left(\lambda,\theta\right)\equiv\sqrt{1+\lambda^{2}-2m_{2}\lambda\cos\left(\theta\right)}$.
Note that for $\phi=0$ (spatially homogeneous field) $\gamma_{G}^{n}$
does not depend on $t_{LR}$. This fact, implies that it remains invariant
for several tunnel coupled two level systems as long as $\phi=0$
(i.e. the field is spatially homogeneous). Once we add the spatial
anisotropy through the phase difference $\phi\neq0$, Berry's result
is modified. It adds an interplay between tunneling and ac-field which
has not been previously analyzed.

The next step is the calculation of the curvature 2-form $F^{n}=dA^{n}=\left(\frac{\partial}{\partial\varphi}A_{\theta}^{n}-\frac{\partial}{\partial\theta}A_{\varphi}^{n}\right)d\varphi\wedge d\theta$,
that will be used for the calculation of the Chern numbers:
\begin{eqnarray}
F_{\phi=0}^{n} & = & m_{1}\frac{\sin\left(\theta\right)}{2}\label{eq:Curvature1}\\
F_{\phi=\pi}^{n} & = & m_{1}\frac{\sin\left(\theta\right)}{2}\frac{1-\lambda m_{2}\cos\left(\theta\right)}{\left(1+\lambda^{2}-2\lambda m_{2}\cos\left(\theta\right)\right)^{3/2}}.\label{eq:Curvature2}
\end{eqnarray}
The result shows that for $\phi=\pi$, the original curvature (Eq.\ref{eq:Curvature1})
is renormalized by a factor which depends on $t_{LR}$ and $m_{2}$
(Eq.\ref{eq:Curvature2}). For $\lambda=1$, the curvature 2-form
inverts its value, and at the limit of large tunneling $t_{LR}\gg B$
the curvature tends to zero for all $\theta$, evolving towards a
flat space. The first Chern number is given by:
\begin{equation}
c_{1}^{n}\left(\lambda\right)=\int_{\mathcal{S}^{2}}\frac{F^{n}}{2\pi}=\begin{cases}
m_{1}\Theta\left(1-\lambda\right) & \text{for\ \ensuremath{\phi}=\ensuremath{\pi}}\\
m_{1} & \text{for\ \ensuremath{\phi}=0}
\end{cases}\label{eq:Chern1}
\end{equation}
being $\Theta\left(x\right)$ the Heaviside step function. Hence,
for $\phi=\pi$, the system undergoes a transition from a topological
phase $\mathbb{Z}$ ($c_{1}\neq0$) to a trivial phase ($c_{1}=0$)
at $\lambda=1$ by tuning the ratio between the hopping $t_{LR}$
and the intensity of the ac-field $|\vec{B}|$.

All results in this section has been obtained under the adiabatic
assumption, but this is an approximation for ideal slow evolution.
Therefore a natural question arises: How do the previous results change
when we consider non-adiabatic evolution?.

\paragraph*{Non-adiabatic case:}

Aharonov and Anandan (A-A) proposed a generalization of the Berry
phase to non-adiabatic processes, the A-A phase $\gamma_{\text{A-A}}$
\citep{Anandan87}. As later was pointed out, this generalization
corresponds to a universal connection $\mathcal{A}$ in a universal
PFB $\mathbb{C}P^{\infty}$ or A-A bundle\citep{CPcomment}. It establishes
a general relation between the classification of $U\left(1\right)$
PFB and quantum mechanical systems\citep{Bohm1993}. The relation
between Berry's and Anadan's result is stablished through a function
$f_{n}^{*}$ (pullback bundle), that maps curves $\mathcal{C}$ defined
in $\mathbb{C}P^{\infty}$ to curves $C$ in the parameter space,
i.e. $C=f_{n}^{*}\left(\mathcal{C}\right)$, then $f_{n}^{*}$ has
all the information for the computation of the geometric phase (the
explicit form of the Hamiltonian determines $f_{n}^{*}$).

We will show below that for the non-adiabatic case, the geometric
phase $\gamma_{\text{A-A}}$ and the phase diagram match the previous
results in the adiabatic limit\citep{Geometrycomment}.

Considering the present setup in the non-adiabatic regime, we shall
demonstrate that it can be exactly solved (without the adiabatic approximation),
and calculate the geometric phase $\gamma_{\text{A-A}}$ and the topological
invariants.

By means of a unitary transformation to a co-rotating frame with the
ac-field, we obtain a static Hamiltonian which can be diagonalized.
The unitary transformation is given by $U\left(t\right)=\exp\left\{ -i\Omega t\left(S_{z}^{L}+S_{z}^{R}\right)\right\} $,
and the transformed Hamiltonian $\tilde{H}=U^{\dagger}HU-iU^{\dagger}\dot{U}$
reads:
\begin{eqnarray}
\tilde{H} & = & \sum_{i=L,R}\left(B\cos\left(\theta\right)-\Omega\right)S_{z}^{i}+\sum_{\sigma,i\neq j}t_{LR}c_{\sigma,i}^{\dagger}c_{\sigma,j}\nonumber \\
 &  & +\sum_{i=L,R}B\sin\left(\theta\right)\left[\cos\left(\phi_{i}\right)S_{x}^{i}+\sin\left(\phi_{i}\right)S_{y}^{i}\right].\label{eq:Transformed-Ham}
\end{eqnarray}
which is independent of $t$. Note that the time dependence of the
transformation renormalizes the energy levels by $\Omega$.

The diagonalization of the Hamiltonian leads to the energies $\mathcal{E}_{n}\left(\phi\right)$:
\begin{eqnarray}
\mathcal{E}_{n}\left(0\right) & = & -m_{1}\frac{B}{2}\sqrt{1+\mu^{2}-2\mu\cos\left(\theta\right)}-m_{2}t_{LR}\nonumber \\
\mathcal{E}_{n}\left(\pi\right) & = & -m_{1}\frac{B}{2}\sqrt{1+\Delta_{m_{2}}^{2}-2\Delta_{m_{2}}\cos\left(\theta\right)}\label{eq:Energy2}
\end{eqnarray}
being $\mu\equiv\Omega/B$ and $\Delta_{m_{2}}\equiv\left(\Omega+2m_{2}t_{LR}\right)/B$.
Note that in both cases, the limit $\Omega\rightarrow0$ matches the
adiabatic result. Note also the similarities between Eq.\ref{eq:Energy1}
and Eq.\ref{eq:Energy2}, where the frequency renormalizes the different
parameters.

Interestingly, the energies present degeneracy points for both $\phi=0,\pi$.
Our calculations show that the geometric phases are also Abelian at
degeneracy points due to the structure of the Floquet operator, which
is in agreement with the results of \citep{Mostafazadeh98}.

The solutions to the Schrdinger equation are given by the transformation
of the eigenvectors $|\tilde{\psi}\rangle$ obtained by diagonalizing
Eq.\ref{eq:Transformed-Ham} to the original frame, i.e. $|\psi\left(t\right)\rangle=U\left(\varphi\left(t\right)\right)|\tilde{\psi}\rangle$.
Therefore, the calculation of the 1-form $\mathcal{A}_{\varphi}$
gives:
\begin{eqnarray*}
\mathcal{A}_{\varphi} & = & i\langle\psi\left(t\right)|\partial_{\varphi}|\psi\left(t\right)\rangle d\varphi=\langle\tilde{\psi}|S_{z}|\tilde{\psi}\rangle d\varphi,
\end{eqnarray*}
being the A-A geometric phase $\gamma_{\text{A-A}}=\oint_{\mathcal{C}}\mathcal{A}_{\varphi}=2\pi\mathcal{A}_{\varphi}$,
i.e. :
\begin{eqnarray}
\gamma_{\text{A-A}}^{n}\left(0\right) & = & m_{1}\pi\frac{\left(\mu-\cos\left(\theta\right)\right)}{\sqrt{1+\mu^{2}-2\mu\cos\left(\theta\right)}}\label{eq:Geometricphase-NonAd1}\\
\gamma_{\text{A-A}}^{n}\left(\pi\right) & = & m_{1}\pi\frac{\left(\Delta_{m_{2}}-\cos\left(\theta\right)\right)}{\sqrt{1+\Delta_{m_{2}}^{2}-2\Delta_{m_{2}}\cos\left(\theta\right)}}\label{eq:Geometricphase-NonAd2}
\end{eqnarray}
and the curvature:
\begin{eqnarray}
F^{n}\left(0\right) & = & m_{1}\frac{\sin\left(\theta\right)}{2}\frac{1-\mu\cos\left(\theta\right)}{\left(1+\mu^{2}-2\mu\cos\left(\theta\right)\right)^{3/2}}\label{eq:Curv-NA}\\
F^{n}\left(\pi\right) & = & m_{1}\frac{\sin\left(\theta\right)}{2}\frac{\left(1-\Delta_{m_{2}}\cos\left(\theta\right)\right)}{\left(1+\Delta_{m_{2}}^{2}-2\Delta_{m_{2}}\cos\left(\theta\right)\right)^{3/2}}\nonumber 
\end{eqnarray}
We can see that the geometric properties for $\phi=0$ does not depend
on $t_{LR}$ nor on $m_{2}$ (always a degeneracy for states with
different $m_{2}$ is present). By contrary, the $\phi=\pi$ case
presents different curvature for each state, but because of the functions
$\Delta_{\pm}$ cannot be independently modified by changing the system
parameters $B$, $\Omega$ and $t_{LR}$, the geometric phases and
curvatures also cannot. The Chern numbers can be calculated through
the curvature, and then, obtain the phase diagrams as a function of
$\Delta_{m_{2}}$ or $\mu$:
\begin{eqnarray}
c_{1}^{n}\left(0\right) & = & \frac{m_{1}}{2}\left(1+\text{sign}\left(1-\mu\right)\right)\label{eq:Chern2}\\
c_{1}^{n}\left(\pi\right) & = & m_{1}\Theta\left(1-\left|\Delta_{m_{2}}\right|\right)\nonumber \\
 & = & m_{1}\Theta\left(1-\left|\Omega+2m_{2}t_{LR}\right|/B\right)\nonumber 
\end{eqnarray}
The results obtained for the geometric phase show certain interesting
features. For the $\phi=0$ case, the non-adiabatic phases $\gamma_{\text{A-A}}$
present some corrections in $\mu$ (which is precisely the adiabatic
parameter) to the adiabatic case, but they remain degenerate in pairs,
and the variation of $\mu$ modifies all geometric phases simultaneously.
Also we must note that $\gamma_{\text{A-A}}^{n}\left(\phi=0\right)$
(Eq.\ref{eq:Geometricphase-NonAd1}), as in the adiabatic case (Eq.\ref{eq:BerryphaseAd}),
does not depend on $m_{2}$ nor on $t_{LR}$, meaning that systems
with a larger number of coupled spins would have the same geometric
phase.

Regarding the phase diagram, the $\phi=0$ case just contains two
phases: The topologically trivial and the $\mathbb{Z}$ phase. The
trivial phase appears when we go out of the adiabatic regime by increasing
the frequency ($\mu=\Omega/B>1$), while the $\mathbb{Z}$ phase governs
the whole adiabatic regime.

\begin{figure}[h]
\includegraphics{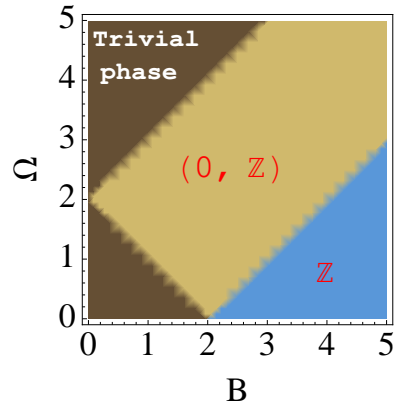}

\caption{\label{fig:Phase-diagram2}Topological phases diagram vs $\left(B,\Omega\right)$
for fixed $t_{LR}=1$ and $\phi=\pi$. The phase diagram shows the
physically accessible topological regions. Note that the adiabatic
limit is found by choosing $\Omega=0$ with the trivial and fully
topological phases described above. In addition, for non adiabatic
processes (i.e. $\Omega>0$) a new phase $\left(0,\mathbb{Z}\right)$
appears, in which just half of the states are characterized by a non
vanishing Chern number.}
\end{figure}

Concerning the $\phi=\pi$ case in the non-adiabatic regime, the geometric
phases (Eq.\ref{eq:Geometricphase-NonAd2}) depend on two functions
$\Delta_{m_{2}}$ , and hence on both indexes $\left(m_{1},m_{2}\right)$
of the state vector. This allows to modify the geometric and topological
properties differently for each state. This is an important difference
with the $\phi=0$ case, because now the geometric phase for each
state behaves differently when we vary $t_{LR}$, $\Omega$ and $B$
for $\phi=\pi$. This is reflected in the phase diagram, where two
new topological phases appear: $\left(0,\mathbb{Z}\right)$ and $\left(\mathbb{Z},0\right)$
(first index labels states with $m_{2}=+1$ and second index states
with $m_{2}=-1$) in which only two of the states have non-vanishing
Chern number. More interestingly, at arbitrary high frequencies $\Omega$
we show that states with non-vanishing Chern number can appear (Fig.\ref{fig:Phase-diagram2}).

Because the functions $\Delta_{\pm}$ depend both on the system parameters,
we plot the phase diagram as a function of $B$ and $\Omega$ for
fixed $t_{LR}$. Interestingly, the $\left(\mathbb{Z},0\right)$ phase
is not accessible (Fig.\ref{fig:Phase-diagram2}). This is due to
dependence on $B$, $t_{LR}$ and $\Omega$ of the functions $\Delta_{\pm}$.
Note that for a different dependence on the parameters, the $\left(\mathbb{Z},0\right)$
phase could appear. We also observed that the variation of the parameter
$t_{LR}$ only modifies the width of the $\left(0,\mathbb{Z}\right)$
phase, such that in the limit $t_{LR}\rightarrow0$ we obtain the
phase diagram for the $\phi=0$ case.

The existence of phases with non-vanishing Chern number implies a
non-trivial bundle, and therfore holonomy elements different from
the identity. It means that it is impossible to avoid the geometric
phases in the study of these systems. The geometric phases are local,
in difference with the Chern numbers, and can be different from zero
even for regions where $c_{1}=0$. Our results can be important for
experiments involving the measurement of geometric phases, which can
be performed by using superconducting qubits and quantum state tomography\citep{Leek07}
among other setups\citep{Nagasawa12}.

\paragraph*{Summary:}

Summarizing, we analyzed the geometric phases and topology of tunnel
coupled systems with an on-site pseudo-spin degree of freedom, and
driven by ac-fields. We have considered the case of an electron spin,
tunneling between two different sites, which is coupled to an ac magnetic
field that produces transitions between the spin up/down levels. The
interplay between the spatial and spin degree of freedom due to the
ac magnetic field, as well as its dependence with the external field
parameters $B$, $\Omega$ and $\phi$ (phase difference between different
sites) is analyzed from a geometrical point of view.

Interestingly, we found that a system presenting spatial anisotropy
(in our case due to the phase difference $\phi$ between the external
ac driving fields in different sites) leads to a complex and rich
behavior compared with Berry\textquoteright{}s classical result for
a localized spin in the adiabatic regime. Our analysis is also extended
to the non-adiabatic regime. In this case we find geometrical phases
which depend in a non trivial way on the spatial field anisotropy.
The results generalize the well known Aharonov-Anandan phase for localized
spins, resulting in a novel topological phase diagram (Fig.\ref{fig:Phase-diagram2}).
Our results are the basis for the analysis of larger systems with
spatial periodicity which will be the subject of a future work.

Although the topological phase transition can be difficult to be directly
measured in our setup, the measurement of geometric phases could indirectly
show the topological phase transition (see Fig.\ref{fig:Geometric-phase-variation}).
Setups involving quantum dots in slanting magnetic fields \citep{Pioro2008}
and quantum circuits\citep{Leek07} can be used for this purpose.
Also, the non-adiabatic analysis shows how the topological invariants
change out of the adiabatic regime, leading to possible applications
in non-adiabatic quantum computation.

We acknowledge MAT 2011-24331 and ITN, grant 234970 (EU) for financial
support. A. G\'{o}mez-Le\'{o}n acknowledges JAE program.

\bibliographystyle{apsrev4-1}
\bibliography{TopologyRef}

\begin{thebibliography}{24}
\bibitem [{CPc()}]{CPcomment}%
  \BibitemOpen
  \href@noop {} {\bibinfo  {journal} {$\mathbb{C}P^{N}$ is the complex
  projective space of N dimension. This is the set of equivalence classes P(H)
  for states $| \psi \rangle \in H$, such that the equivalence relation is
  given by $| \psi \rangle \sim c | \psi \rangle$, being H the Hilbert space of
  dimension N and c a complex number}\ }\BibitemShut {NoStop}%
\bibitem [{Geo()}]{Geometrycomment}%
  \BibitemOpen
\bibfield  {journal} {  }\href@noop {} {\bibinfo  {journal} {Note that the
  Aharonov-Anandan PFB is a fully geometric entity and does not assume any
  approximation such us adiabaticity}\ }\BibitemShut {NoStop}%
\end{thebibliography}

\end{document}